



\def\NI{\noindent}

\long\def\UN#1{$\underline{{\vphantom{\hbox{#1}}}\smash{\hbox{#1}}}$}
\magnification=\magstep 1
\overfullrule=0pt
\hfuzz=16pt
\voffset=0.0 true in
\vsize=8.8 true in
   \def\NP{\vfil\eject}
   \baselineskip 20pt
   \parskip 6pt
   \hoffset=0.1 true in
   \hsize=6.3 true in
\nopagenumbers
\pageno=1
\footline={\hfil -- {\folio} -- \hfil}

\centerline{\UN{\bf Locally Frozen Defects in Random Sequential}}

\centerline{\UN{\bf Adsorption with Diffusional Relaxation}}

\vskip 0.4in

\centerline{\bf Jian--Sheng Wang{\rm ,}$^{a,b}$\ Peter Nielaba$^c$\
{\rm and}\ Vladimir Privman$^d$}

\vskip 0.2in

\NI\hang $^a${\sl Department of Physics, Hong Kong Baptist College,
224 Waterloo Road, Kowloon, Hong Kong}

\NI\hang $^b${\sl Present address: c/o Department of Physics,
National University of Singapore, Kent Ridge, Singapore 0511}

\NI\hang $^c${\sl Institut f\"ur Physik, Universit\"at Mainz,
Staudingerweg 7, D--55099 Mainz, Germany}

\NI\hang $^d${\sl Department of Physics, Clarkson University,
Potsdam, New York 13699--5820, USA}

\vskip 0.4in

\centerline{\bf ABSTRACT}

Random sequential adsorption  with diffusional relaxation, of
two by two square objects on the two-dimensional square lattice is
studied by Monte Carlo computer simulation. Asymptotically for
large lattice sizes, diffusional relaxation allows the deposition
process to reach full coverage. The coverage
approaches the full occupation value, 1, as a power-law
with convergence exponent near $1 \over 2$. For a periodic lattice
of finite (even) size $L$, the final state is a frozen random
rectangular grid of domain walls connecting single-site defects.
The domain sizes saturate at $\sim L^{0.8}$. Prior to saturation,
i.e., asymptotically for infinite lattice, the domain growth is
power-law with growth exponent near, or possibly somewhat smaller
than, $1 \over 2$.

\vfill

\NI {\bf PACS numbers:}$\;$  68.10.Jy, 02.50.+s, 82.65.-i

\NP

\NI \UN{\bf 1.~Introduction}

\

Random sequential adsorption (RSA) models have been studied
extensively due to their relevance to surface deposition [1-2]. The
depositing particles are represented by hard-core extended objects;
they are not allowed to overlap. In monolayer deposition of
colloidal particles and macromolecules [3-6] one can further assume
that the adhesion process is irreversible. However, recent
experiments on protein adhesion at surfaces [7-8] indicate that in
biomolecular systems effects of surface relaxation, due to
diffusional rearrangement of particles, are observable on time
scales of the deposition process. The resulting large-time coverage
is denser than in irreversible RSA and in fact it is experimentally
comparable to the fully packed (i.e., locally semi-crystalline)
particle arrangement.

Irreversible RSA has been studied extensively by many authors
[1-2]. The most interesting aspect of such processes is the
power-law large-time convergence to the jamming coverage in
continuum off-lattice deposition. This slow time-dependence, as
opposed to exponential convergence in lattice deposition, is due to
gaps arbitrary close in size (and shape) to that of the depositing
particles and therefore reached with low probability. Asymptotic
arguments describe this rare-event dominated process [9-10].
Crossover from lattice to continuum can be also elucidated
analytically [11].

Studies of RSA with added diffusional relaxation by
analytical means encounter several difficulties associated with
collective effects in hard-core particle systems at high
densities (such as, for instance, phase separation), and with the
possibility, in certain lattice models, of locally ``gridlocked''
vacant sites.

Both complications are not present in $1d$: there are no
equilibrium phase transitions (in models without deposition), traces
of which might manifest themselves as collective effects in $d>1$
RSA with diffusion [12], and furthermore diffusional
relaxation leads to simple hopping-diffusion interpretation of the
motion of vacant sites in $1d$ which recombine to form larger open
voids accessible to deposition attempts. Thus, both extensive
numerical studies and their analytical interpretation were possible
in $1d$ [13-15].

For $d>1$ models, a low-density-expansion approximation scheme was
applied [16] to off-lattice deposition of circles on a plane,
accompanied by diffusional relaxation. However, no
analytical studies were reported of the high-density behavior and
the associated collective effects.

Extensive numerical simulations were reported [12], of the
RSA process with diffusional relaxation, for the lattice
hard-square model [17], i.e., the square-lattice hard-core model
with nearest-neighbor exclusion. This model is well studied for its
equilibrium phase transition [17] which is second-order with
disordered phase at low densities and two coexisting ordered
phases, corresponding to two different sublattice particle
arrangements, at high densities. Another simplifying feature of
the hard-square model is that the only possible gridlocked
(locally frozen) vacancies are parts of domain walls [12]. As a
result the coverage reaches the full crystalline limit at large
times, by a process of diffusional domain wall motion leading to
cluster growth reminiscent of spinodal decomposition in quenched
binary alloys and fluids at low temperatures [18-20].

In this work we report extensive computer simulation results
for RSA with diffusion on the two-dimensional square lattice
with objects occupying two by two squares of four sites. One can
view the deposition of such $2\times 2$ objects as equivalent to
a hard-core model with nearest-neighbor and
next-nearest-neighbor exclusions. The distinctive feature of this
model is the existence of locally frozen single-site defects. RSA
with diffusion on a periodic lattice then leads to frozen states
with domains of four different phases.  The corresponding
equilibrium ground states are highly degenerate [17,21-22].

The model is detailed in Section~2. The frozen states in the
large-time limit are described in Section~3. Time-dependence of
the single-site defect density, of the coverage, and of the
ordered domain size, are discussed in Sections~4, 5, 6, respectively.
These sections also contain discussion of the results.

\

\

\NI \UN{\bf 2.~Definition of the Model}

\

We consider a square lattice of size $L \times L$, where $L$ is
even. Initially all sites are empty. Each trial Monte Carlo step
starts with choosing at random a lattice site
$(i,j)$. With probability $p$ we try deposition of
a two by two square to cover sites $(i,j)$, $(i+1,j)$, $(i,j+1)$,
$(i+1,j+1)$.  Deposition attempt is successful if these four sites
are all empty. The sites are then marked as occupied. In the
actual computer program the site $(i,j)$ is further listed as
the lower-left corner of a block, for later use.

We perform
a diffusion attempt with probability $1-p$. Diffusional move is
attempted only if the chosen site $(i,j)$ is already one of the
lower-left
corner sites. We try to diffuse (move) the square in one of the four
directions (up, down, left, right) chosen at random, by one lattice
spacing.  Diffusion is successful if the move is not blocked by
other squares. The time $t$ is measured in terms of Monte Carlo
steps per site: one time unit is defined as $L^2$ deposition
or diffusion attempts (successful or not).

In comparing data for different $p$ values, it is further
convenient to rescale time by $p$, see [12-14] for details.
However, we report specific results only for $p=0.1$ here.
Furthermore, all lattice sizes used were powers of 2. This
restriction to one $p$ value, and generally the scope of the
results presented, were due to numerical difficulties involved
in simulating RSA with diffusion in $2d$. For instance, results
reported here took over 2 months of CPU time on an HP Apollo Model 720
workstation. However, we checked some of the observations reported
in the later sections also for $p=0.8$. The formation of the
frozen states on periodic even-sized lattices (Section~3) was also
checked for sizes other than powers of 2, and for several $p$
values other than $0.1$.

The corresponding equilibrium model has been studied in [17, 21-22].
The equilibrium phase transition from the low-density disordered
state to the high-density ordered state with four phases, is still
not fully classified. The distinctive feature of the equilibrium
case is large entropy of the ordered arrangements of $2\times 2$
square objects [17, 21-22].

\

\

\NI \UN{\bf 3.~Frozen States at Large Times}

\

Let us consider first the case of the periodic (even-sized)
two-dimensional square lattice. Due to diffusional relaxation
the coverage $\theta$ reaches almost the fully crystalline value
1 for large times, $t \to \infty$.  However, the final
configuration on a finite-size lattice typically has frozen
defects which are single-site and serve as points of origin
of domain walls separating four different sublattice
arrangements (``phases''). One such configuration is illustrated
in Figure~1. The defects are locally ``gridlocked,'' and
any state that contains all the empty area in such
single-site isolated defects no longer evolves
dynamically.

The defect lines are either horizontal or vertical so that the
resulting structure is essentially a random rectangular grid. There
are no other types of defects here that can survive in the
large-time limit, unlike other studies of the formation of
two-dimensional frozen polycrystalline structures (by other growth
mechanisms) available in the literature [23-24].

This freezing-in rectangular grid was also observed for a related
equilibrium model [21] by slow cooling from a disordered
configuration to zero temperature. The frozen network seems to
depend profoundly on the boundary conditions; the rectangular grid
is characteristic of periodic boundary conditions. Indeed, when a
fixed boundary is used, where the particles are not allowed to
occupy the sites outside the $L \times L$ square ($L$-even), the
final configuration is always a regular pattern of $L\times L$
sites. In this case, it is obvious that whenever there are
unoccupied defect sites, there must be also some unoccupied sites at
the boundary. Through diffusion and deposition, the defect sites in
the interior as well as the empty sites at the boundary are fully
eliminated.

The geometrical nature of the frozen grid pattern is illustrated
by the following consideration. On a
periodic two-dimensional lattice with even dimensions, the
``frozen'' network
will have the geometry of the rectangular grid (Figure~1), i.e., a
distribution of rectangular fully ordered
domain shapes. Since each horizontal domain wall implies a
shift of 1 lattice site, on an even periodic lattice the number of
horizontal domain walls in each cross-section must be even.
Similarly the number of vertical domain walls must be even. Thus
the total number of domains, equal the number of single-site
defects, is a product of two even numbers and must therefore be a
multiple of 4. On the fixed-boundary lattice defined earlier, one
simply cannot fit a rectangular array of domain walls with all
vertices of coordination number 4. Thus, frozen states are not
allowed due to purely geometrical constraints.

Let $\rho (t)$ denote the density of the single-site defects.
Numerically, we counted the number of
unoccupied single sites fully surrounded by occupied sites,
and divided by the total number of lattice sites, $L^2$. The
large-time limiting values are plotted in Figure~2 vs.~the linear
system size $L$.  For system sizes $L>128$ it is difficult
to reach the final limiting state. The data shown were
obtained for the largest simulation times and they may be just upper
bounds on the actual asymptotic values.

The data indicate that the
density decreases at least as fast as $1/L$. This also yields the
deviation of the coverage at large time from the monocrystalline
value 1. Thus, $1-\theta (\infty) \sim 1/L$, on a periodic lattice.
The number of domains in the rectangular grid is equal to
the number of vertices. If the domain size distribution were
sharply peaked at some central value then the domain areas would
saturate at $\sim L$. However, numerical evidence presented
in Section~6 suggests that the domain linear size squared rather
grows $\sim L^{1.6}$. This seems to suggest that the domain
size, and possibly shape (i.e., rectangle size ratio), distributions
are nontrivial and must be further investigated. Accurate domain
statistics would require simulation of much larger lattices
then those reported here.

\

\

\NI \UN{\bf 4.~Time-Dependence of the Single-Site Defect Density}

\

The single-site density $\rho (t)$ is
plotted in Figure~3, on logarithmic scale, for several system
sizes, $L$. It is interesting to note that $\rho (t)$ has a peak
around $t=10$. For times $t<10$ the single-site
density increases with time, and then decreases for large
$t$.  This is presumably due to the difference between
dominant dynamical mechanisms for short and long times. In fact,
results of this and later sections suggest that the dynamics has
three distinct regimes.

\NI\hang \UN{\sl Regime I.}\ \  For short times, the
configuration is mainly build up by deposition. All types of
structures, including the single-site defects, are produced.

\NI\hang \UN{\sl Regime II.}\ \  At later times cluster
coarsening occurs leading to ordering and formation of the
frozen-grid defect structure. The defect structure in turn evolves
from complicated to grid-like.

\NI\hang \UN{\sl Regime III.}\ \  The latest stage of the dynamics
consists of straightening up of defect lines and final freezing of
the rectangular grid. This regime will be further commented on in
the later sections.

Finite-size effects show up already in Regime II: $\rho (t)$
saturates at $\sim 1/L$. The curves in Figure~3 then break off
the main ``envelope'' which presumably corresponds to domain
coarsening before the finite-size effects set in. This infinite-$L$
envelope in Regime II, i.e., for times $t>50$ in Figure~3, fits
quite accurately the power law

$$ \rho(t) \sim t^{-0.573 \pm 0.004}\;\; . \eqno(1) $$

\NI The precision of the power-law fit here is much higher than
all other such fits reported in the present work. This presumably
reflects absence of competing mechanisms in the
Regime II.

However, the exponent value for the growth of the linear
domain size assuming one domain per one single-site defect,
half the value in (1), $0.2865 \pm 0.002$, is smaller than the
accepted value, 0.5, for the non-conserved dynamics
in $2d$ [18-20]. Thus the simple domain-per-defect picture does
not hold suggesting that Regime II corresponds not just to
cluster coarsening but also to rearrangement of the empty area which
consists not only of isolated single-site defects but
includes more complicated, evolving structures. Thus, the power law
in (1) cannot be simply explained by theories of cluster
coarsening. The latter process will be further addressed in
Section~6.

\

\

\NI \UN{\bf 5.~Time-Dependence of the Coverage}

\

The empty area fraction, $1-\theta$, is shown in Figure~4.
Generally for large times we expect the empty area to be
represented by terms with different time dependence: single-site,
dimer vacancies, trimers, etc.,

$$ \theta(t) = 1 - \rho(t) - 2 \rho_{\rm dimer}(t) - \ldots \;\; .
\eqno(2)$$

\NI Specifically, the eventual saturation of the different-$L$
curves suggested in Figure~4, is due to the fact that $\rho (t)$
remains small but finite (of order $1/L$) for large times.

Consideration of the actual dynamical time steps achieved by writing
a ``movie'' program to reproduce the consecutive configurations on
the X-windows screen, has lead us to believe that the
longest-surviving multiple-site empty area fragments are dimers.
In the time Regime III, the grid structure has already been formed.
However, the domain walls have kinks which are formed by dimer
(two-site) empty-area defects. These can diffuse along the domain
walls and ``interact'' with single-site defects while turning
90$^\circ$ at each such encounter. The kinks can also sometimes
``annihilate'' pairwise due to deposition when they come in
contact in the appropriate configurations.

A mapping to a reaction-diffusion system is suggested. However,
here it will be more complicated than in $1d$ [13-15] where the
``diffusers'' were all identical and annihilated (with some
probability) on pairwise encounters. If we speculate that the
random grid is effectively one-dimensional and the
reaction-diffusion system of dimer defects is otherwise similar to
the $1d$ case, then we have the prediction

$$ \rho_{\rm dimer}(t) \sim 1/ \sqrt{t} \;\; , \eqno(3) $$

\NI for large times, in the Regime III.

Upon onset of saturation, the curves in Figure~4 become
size-dependent. If we select the portions of the curves which
follow the $L$-independent ``envelope,'' then we can fit the
effective power-law,

$$ 1 - \theta(t) \sim t^{-\alpha} \; \; . \eqno(4) $$

\NI However, least-square fits give exponent $\alpha$ values
decreasing from 0.61 to 0.53, as $t$ increases.
This deviation from the prediction (3) may
be in part due to the use of 1 instead of the correct,
$L$-dependent limiting value $\theta( \infty ) = 1 - \rho ( \infty
) $, and in part due to higher-order terms in (2). Indeed, for
lattice sizes and times reached in our simulation the effective
power-law fit here can be seen to largely cover Regime II rather
than Regime III.

\

\

\NI \UN{\bf 6.~Domain Size Measures}

\

We considered two definitions for the linear domain
size. The first is obtained via the average
``magnetization'' vector, ${\bf M}$, where each $2 \times 2$
object  is labeled by a unit vector,
directed up, down, left, right depending on
which sublattice the object is in.  The domain size is defined by

$$ \ell_m = {4\over L } \left\langle {\bf M}^2 \right\rangle^{1/2} \;\; .
\eqno(5) $$

\NI Here ${\bf M}$ is defined as the vector sum of the unit
vectors of each of the deposited objects. The average $\langle
\ldots \rangle$ is with respect to different Monte Carlo runs. Note
that for a uniform single domain without defects, the domain
size is $L$.

The second definition is via the average square cluster size,

$$\ell_c =  {4\over L } \bigl\langle \sum_i c_i^2 \bigr\rangle^{1/2}
\;\; , \eqno(6) $$

\NI where the clusters are defined as sets of
squares on the same sublattice and continuously connected with
each other; $c_i$ is the number of squares in the cluster $i$.

Figure~5 shows the domain sizes, $\ell_c$, vs.~time, for various values
of $L$.  The results for $\ell_m$ (not shown) is similar but less
accurate.  By comparison between different sizes, we concluded
that the data for $L=512$ and $t<2000$ should not show finite-size
effect. Power-law least-squares fit

$$ \ell_{m,c} \propto t^\beta\;\;  \eqno(7)$$

\NI to the infinite-size result gives the exponent $\beta$
varying from 0.50 (for small $t$) to 0.36
(for large $t$) in different time intervals.
It is therefore unlikely that the standard
non-conserved-dynamics cluster growth exponent, $1 \over 2$
[12,18-20], can be accurately confirmed for this model without
further much more extensive simulations.

The domain size saturates for finite $L$.  The dependence of the
saturation value on the size $L$ is plotted in Figure~6, on
logarithmic scale.  These data are quite well fit by the power law,

$$ \ell_{c,m}(t \to \infty) \sim L^{0.80\pm0.02} \;\; . \eqno(8)$$

Finally, let us stress that the results of this section were for
periodic boundary conditions. Preliminary data indicate that the
behavior of the cluster sizes, at least as far as their magnitudes
go, is quite different for fixed boundary conditions.

In summary we reported empirical observations and numerical exponent
fits for the problem of lattice RSA with diffusional relaxation
which is distinctive in that isolated defects are frozen in,
leading to domain formation and strong dependence on boundary
conditions. Unfortunately, both analytical understanding of such
dynamical systems is very limited, and numerical results should
be considered as preliminary. Future effort must be focussed
on exploring other models but also on obtaining further
high-quality numerical Monte Carlo results for the $2 \times 2$
model on the square lattice considered in this work.

\

\

\NI \UN{\bf Acknowledgements}

\

This work was supported in part by a Hong Kong Baptist College
Faculty Research Grant (FRG/91-92/II-25) and by the
DFG-Sonder\-forschungs\-bereich 262.
P.N. wishes to thank the DFG for Heisenberg followship.

\NP

\centerline{\bf REFERENCES}

{\frenchspacing

\item{[1]} Review: M.C.~Bartelt and V.~Privman, Int.
J.~Mod. Phys. B{\bf 5}, 2883 (1991).

\item{[2]} Review: J.W.~Evans, Rev. Mod. Phys.
(1993), in print.

\item{[3]} J.~Feder and I.~Giaever, J.~Colloid Interface Sci. {\bf
78}, 144 (1980).

\item{[4]} G.Y.~Onoda and E.G.~Liniger, Phys. Rev. A{\bf
33}, 715 (1986).

\item{[5]} N.~Ryde, N.~Kallay and E.~Matijevi\'c, J.~Chem.
Soc. Farad. Tr. {\bf 87}, 1377 (1991).

\item{[6]} N.~Ryde, H.~Kihira and E.~Matijevi\'c, J.~Colloid
Interface Sci. {\bf 151}, 421 (1992).

\item{[7]} J.J.~Ramsden, preprint (1992).

\item{[8]} See also J.J.~Ramsden, J.~Phys. Chem.
{\bf 96}, 3388 (1992).

\item{[9]} Y. Pomeau, J. Phys. A{\bf 13}, L193
(1980).

\item{[10]} R.H. Swendsen,  Phys. Rev. A{\bf
24}, 504 (1981).

\item{[11]} V.~Privman, J.--S.~Wang and P.~Nielaba,
Phys. Rev. B{\bf 43}, 3366 (1991).

\item{[12]} J.--S.~Wang, P.~Nielaba and V.~Privman, Mod. Phys.
Lett. B{\bf 7}, 189 (1993).

\item{[13]} V.~Privman and P.~Nielaba, Europhys. Lett. {\bf 18}, 673
(1992).

\item{[14]} P.~Nielaba and V.~Privman, Mod. Phys. Lett. B {\bf 6},
533 (1992).

\item{[15]} V.~Privman and M.~Barma, J.~Chem. Phys. {\bf
97}, 6714 (1992).

\item{[16]} G. Tarjus, P. Schaaf and J. Talbot, J.~Chem. Phys. {\bf
93}, 8352 (1990).

\item{[17]} L.K.~Runnels, in {\sl Phase Transitions and Critical
Phenomena}, Vol.~2, p.~305, C.~Domb and M.S.~Green, eds.
(Academic, London, 1972).

\item{[18]} J.D.~Gunton, M.~San~Miguel, P.S.~Sahni,
{\sl Phase Transitions and Critical
Phenomena}, Vol.~8, p.~267, C.~Domb and J.L.~Lebowitz, eds.
(Academic, London, 1983).

\item{[19]} O.G.~Mouritsen, in {\sl Kinetics and
Ordering and Growth at Surfaces}, p.~1, M.G. Lagally, ed. (Plenum,
NY, 1990).

\item{[20]} A.~Sadiq and K.~Binder, J.~Stat. Phys. {\bf 35}, 517
(1984).

\item{[21]} K. Binder and D.P. Landau, Phys. Rev. B{\bf 21}, 1941
(1980).

\item{[22]} W. Kinzel and M. Schick, Phys. Rev. B{\bf 24}, 324
(1981).

\item{[23]} B.D. Lubachevsky and F.H. Stillinger, J. Stat. Phys.
{\bf 60}, 561 (1990), and references therein.

\item{[24]} See also B.D. Lubachevsky, F.H. Stillinger and E.N.
Pinson, J. Stat. Phys. {\bf 64}, 501 (1991).

}

\NP

\centerline{\bf FIGURE CAPTIONS}

\

\noindent\hang{\bf Fig.~1.}\ \ A ``frozen'' configuration with
four defect sites, on the $32 \times 32$ lattice. The $2\times 2$
deposited objects are shown as square outlines while the defects
are shown as four ``holes'' half the linear size of the outlined
objects. The periodic lattice was cut in such a way as to keep the
$2\times 2$ objects whole.

\

\noindent\hang{\bf Fig.~2.}\ \ Density of the
isolated single-site defects at large times, plotted vs.~the linear
system size $L$, on logarithmic scale. The sizes and times,
$(L,\,t)$, were (8, 100), (16, 400), (32,  800),  (64, $10^3$),
(128, $10^4$), (256, $10^4$), (512, $10^5$). The
data were averaged over several hundreds to few thousands of
independent Monte Carlo runs.

\

\noindent\hang{\bf Fig.~3.}\ \ The single-site density plotted
on logarithmic scale as a function of time.
The numbers indicate the system size $L$. The
data were averaged over several hundreds to few thousands of
independent Monte Carlo runs.

\

\noindent\hang{\bf Fig.~4.}\ \ Empty area fraction,
$1-\theta$, plotted vs.~time, $t$, on
logarithmic scale. The system sizes were $L=2^n$,
with $n$ from 3 to 9, from left to right, respectively.

\

\noindent\hang{\bf Fig.~5.}\ \ Effective domain size $\ell_c$
plotted vs.~time $t$ on logarithmic scale, for
system sizes $L$ indicated in the figure.
(The noise in the larger-$L$ data and the break in the $L=64$ data
are, respectively, due to low statistics and due to combining data
from different-statistics runs.)

\

\noindent\hang{\bf Fig.~6.}\ \ Domain size at saturation vs.~the
system size $L$ on logarithmic scale.  The solid circles are data
for $\ell_c$, and open triangles are data for $\ell_m$.  The
straight line has slope 0.8.

\bye